\documentclass[aps,pra,superscriptaddress,floatfix,showpacs,10pt, twocolumn,showkeys]{revtex4}
\usepackage{bbm}
\usepackage{amsfonts}
\usepackage[dvips]{graphicx}
\usepackage{amsmath,amsfonts,amssymb,graphics,graphicx,epsfig,color,times,bbm}

\begin{document}

\title{Multipartite nonlocality swapping}
\author{Chao Li}
\affiliation{Key Laboratory of Opto-electronic Information
Acquisition and Manipulation, Ministry of Education, School of
Physics {\&} Material Science, Anhui University, Hefei 230039,
People's Republic of China.}

\author{Ming Yang \footnote{Corresponding Author: mingyang@ahu.edu.cn}}
\affiliation{Key Laboratory of Opto-electronic Information
Acquisition and Manipulation, Ministry of Education, School of
Physics {\&} Material Science, Anhui University, Hefei 230039,
People's Republic of China.}

\author{Qing Yang}
\affiliation{Key Laboratory of Opto-electronic Information
Acquisition and Manipulation, Ministry of Education, School of
Physics {\&} Material Science, Anhui University, Hefei 230039,
People's Republic of China.}

\author{Zhuo-Liang Cao}
\affiliation{Department of Physics {\&} Electronic Engineering,
Hefei Normal University, Hefei 230061, People's Republic of China.}

\begin{abstract}
Nonlocality swapping of bipartite binary correlated boxes can be
realized by a \emph{coupler} ($\chi$) in nonsignaling models. By
studying the swapping process we find that the previous bipartite
coupler can be applied to the swapping of two multipartite boxes,
and then generate a multipartite box with more users than that of
any of the boxes before swapping. Here quantum bound still appears
in the scheme. The bipartite coupler also can be applied to a hybrid
scheme of generating a multipartite extremal box from many PR
boxes. As the analogue of multipartite entanglement swapping, we
generalize the nonlocality swapping of bipartite binary boxes to
multipartite binary boxes by using a multipartite coupler
$\chi_{N}$, and get the probability of success by connecting the
coupler to the generalized Svetlichny inequality. The multipartite
coupler acting on many multipartite boxes makes multipartite
nonlocality swapping be a more efficient device to manipulate
nonlocality between many users. The results show that Tsirelson's
bound for quantum nonlocality emerges only when two of the $n$ boxes
involved in the coupler process are noisy ones.
\end{abstract}



\maketitle

\section{introduction}
Violation of local realism is a fascinating phenomenon of quantum
mechanics (QM). Since the paradox was presented by Einstein, Podolsky
and Rosen\cite{EPR} in 1935, QM was at one time under suspicion
until Bell\cite{Bell} showed that there existed certain settings for
physical experiments that contradicted 'common sense' views of
reality. Bell's theorem was proved by the experiment\cite{Aspect} in
1982 and other more later. Nowadays, quantum nonlocality is getting
the attention it deserves as another special aspect of the quantum
correlations-entanglement.

To get more insight of quantum mechanics, it makes sense to have a
study on the connection between quantum entanglement and quantum
nonlocality. Like Von Neumann entropy and concurrence\cite{Wootters}
for entanglement, various  criteria were put forward to give a
convenient test of nonlocality. The earliest one-Bell
inequality\cite{Bell} which comes from Bell's theorem gives a
criterion that classic local correlation must obey. Numbers of
Bell-type inequalities\cite{CHSH,CH,Sv} were presented for different
systems afterwards. The violation of Bell-type inequality gives a
straightforward impression about quantum nonlocal correlations
(QNC). A common feature for all these inequalities is that QM's
violation can't reach the maximum value. Naturally, Popescu and
Rohrlich\cite{PR} showed a correlation gives a maximal violation 4
of Clauser-Horne-Shimony-Holt (CHSH) inequality\cite{CHSH}, while
the maximal violation in QM domain, \emph{i.e.} Tsirelson's
bound\cite{bound} $\mathcal{B}_{\mathcal {Q}}=2\sqrt{2}$ by QNC.
Similarly, an algebraic maximum 8\cite{resource} and quantum bound
$4\sqrt{2}$ by Svetlichny inequality\cite{Sv} exist in tripartite
system. With the addition of the correlations over quantum bound,
postquantum correlations (PQC), nonlocal correlations were discussed
in generalized nonsignaling models, as nonsignaling correlated
boxes\cite{resource} decided by joint probability
distribution\cite{nonsign}. Strong information-theoretic
capabilities were revealed in the study of PQC\cite{post}, such as
secure cryptography\cite{secure} and the reduction of communication
complexity\cite{NLDC,complexit}.

Properties of entanglement, such as monogamy, distillation, and
swapping, were proved to be capable for nonlocality in nonsignaling
theory\cite{nonsign,NLD,NLS}, but distinctions still exist between
entanglement and nonlocality. Local operations and classic
communication (LOCC) were essential in entanglement distillation
processes, but nonlocality distillation occurs without CC\cite{NLD}.
Only the operations on the input and output of nonlocal boxes are
crucial to the distillation of nonlocality. Dramatically, the device
operating on the outside of the boxes failed to swap the correlation
between nonlocal boxes\cite{NLfS}. Skrzypczyk \emph{et
al.}\cite{NLD} raised the concept of \emph{genuine boxes} and
\emph{coupler} as analogue of the quantum joint measurement for
nonlocality swapping. Clauser-Horne (CH) inequality\cite{CH} as an
appropriate measure of nonlocality was applied to obtain the
possibility of successful swapping. The fact that only PQC can be
successfully swapped in isotropic resources makes the emergence of
Tsirelson's bound under CH expression. Then, Skrzypczyk and
Brunner\cite{coupler} considered all theoretically possible
couplers, ranging from perfect to minimal couplers, for limited
nonlocality, and quantum bound still appeared in their study.

Zukowski \emph{et al.}\cite{ES} proposed the first entanglement
swapping scheme, and it was soon generalized to the case of
generating a three-particle Greenberger-Horne-Zeilinger (GHZ) state
from three Bell pairs\cite{triES}. Bose \emph{et al.}\cite{multiES}
generalized the procedure of entanglement swapping to the
multiparticle case. A natural question whether multipartite nonlocal
correlations can be swapped raises to us. Exploring the analog of
multiparticle entanglement swapping, we'll try to design a
generalized multipartite coupler for realizing multipartite
nonlocality swapping.

Replacing the CH inequality by CHSH inequality\cite{CHSH} and the
generalized Svetlichny inequality\cite{GSI}, we find that the
existing bipartite coupler provides an adequate dynamical process
for two correlated nonlocal boxes, including multipartite ones. But
this bipartite-coupler-based swapping consumes much resources and
seems less efficient in the case of swapping of multi nonlocal
boxes. Then we will show a generalized coupler, which can be applied
on many multipartite boxes,  makes the swapping of many multipartite
boxes succeed in a more efficient way.

The paper is organized as follows. In Sec.\ref{s2} we briefly review
the quantum nonlocality swapping of two bipartite binary correlated
boxes and show the key points therein. In Sec.\ref{s3}, we define the
general multipartite binary correlated boxes after presenting the
form of the generalized Svetlichny inequality for nonsignal systems.
In Sec.\ref{s4} we generalize the nonlocality swapping on two bipartite
boxes to the case of two multipartite boxes by using the bipartite
coupler $\chi$ and give a discussion on the bound of quantum
nonlocal correlation. In Sec.\ref{s5} we attempt to achieve the
nonlocality swapping on three bipartite boxes and then we show the
multipartite coupler $\chi_{N}$ for nonlocality swapping on
arbitrary number of multipartite boxes. Conclusions and remarks are
presented in Sec.\ref{s6}.

\section{Framework}\label{s2}
\begin{figure}[htbp]
\includegraphics[width=0.45\textwidth]{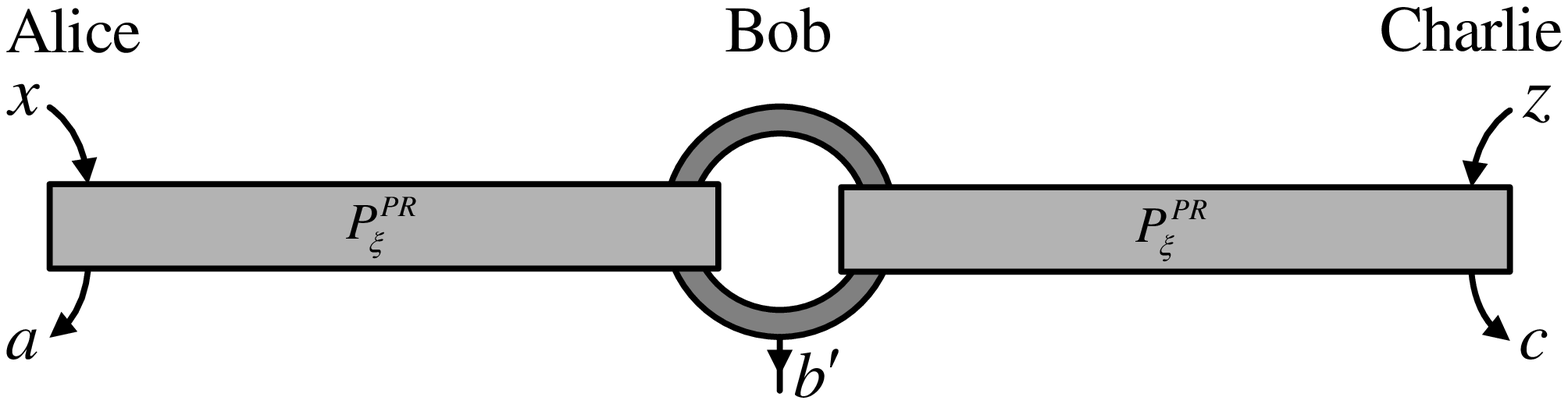}
\caption{\label{22}Nonlocality swapping of two isotropic boxes. The
coupler is the analogue of quantum joint measurement, and quantum
bound (Tsirelson's bound) emerges in this process.}
\end{figure}

Firstly, we'll give a brief review on Skrzypczyk's nonlocality
swapping scheme\cite{NLS}. Bob shares two nonlocal boxes with
another two spatially separated users Alice and Charlie (See
FIG.\ref{22}).

Bob carries out the bipartite coupler $\chi$ on two bipartite
isotropic boxes $P^{\text{PR}}_{\xi}(ab_{1}|xy_{1})$ and
$P^{\text{PR}}_{\xi}(b_{2}c|y_{2}z)$, which are superposition of
Popescu-Rohrlich (PR) box\cite{resource} and the fully mixed box
$\openone$\cite{NLDC}.
\begin{equation}
P^{\text{PR}}_{\xi}={\xi}P^{\text{PR}}+(1-\xi)\openone.
\label{isobox}
\end{equation}
Where $\xi\in[-1,1]$, and when $\xi=-1$, the box is 'anti-PR' box
$P^{\overline{\text{PR}}}$ given by $a\oplus b=xy\oplus1$.

The output $b'=0$ indicates that the correlation between Alice and
Charlie was created successfully. Thus, the whole procedure is
expressed as
\begin{equation}
\begin{aligned}
&P^{\text{PR}}_{\xi}(ab_{1}|xy_{1})P^{\text{PR}}_{\xi}(b_{2}c|y_{2}z)\\
&\xrightarrow{\chi}P(ab'c)=
\begin{cases}
p(b'=0)P^{\text{PR}}_{\xi^{2}}(ac|xz),\\
p(b'=1)P^{f}(ac|xz).
\end{cases}\\
\end{aligned}
\end{equation}

Some characteristics for the swapping process were summarized as
follows:

A coupler working on the genuine part\cite{NLS} of the 'quantum' box
but not on the measurement devices, is different from the action in
Ref.\cite{NLfS}.

The successful probability
$p(b'=0)=\frac{2}{3}\Vec{\text{CH}}\cdot\Vec{P}(b_{1}b_{2}|y_{1}y_{2})\\=\frac{2}{3}[P(11|00)+P(00|10)+P(00|01)-P(00|11)]$
is proportional to the nonlocalities of the boxes which the coupler
is applied on, \emph{i.e.}, the box $P(b_{1}b_{2}|y_{1}y_{2})$'s
violation of CH inequality\cite{CH}. The coupler doesn't always
successfully swap the correlations, with an exception that the
correlation which coupler acted on is PR correlation. Generally,
$P(b_{1}b_{2}|y_{1}y_{2})=\openone$ and the optimal probability of
success is $\frac{1}{3}$.

A successful coupling process is a linear transformation. By this
property, we can get the results as follows immediately:
\begin{subequations}
\begin{equation}
P^{\text{PR}}(ab_{1}|xy_{1})P^{\text{PR}}_{\xi}(b_{2}c|y_{2}z)\xrightarrow{\chi(b'=0)}P^{\text{PR}}_{\xi}(ac|xz),
\end{equation}
\begin{equation}
\openone(ab_{1}|xy_{1})P^{\text{PR}}_{\xi}(b_{2}c|y_{2}z)\xrightarrow{\chi(b'=0)}\openone(ac|xz),
\end{equation}
\begin{equation}
P^{\overline{\text{PR}}}(ab_{1}|xy_{1})P^{\text{PR}}_{\xi}(b_{2}c|y_{2}z)\xrightarrow{\chi(b'=0)}P^{\text{PR}}_{1-\xi}(ac|xz).
\end{equation}
\end{subequations}
Notice thai Eq.(\ref{isobox}) with the coefficient $\xi\in[-1,1]$ is
the correlation to be swapped and $\xi\in[0,1]$ is the allowed
correlations where the coupler works fine.

When $\xi=1$, we got
\begin{equation}
P^{\text{PR}}(ab_{1}|xy_{1})P^{\text{PR}}(b_{2}c|y_{2}z)\xrightarrow{\chi}
\begin{cases}
\frac{1}{3}P^{\text{PR}}(ac|xz)&\text{if }b'=0,\\
\frac{2}{3}P^{f}(ac|xz)&\text{if }b'=1.
\end{cases}
\end{equation}
Coincidently, the coupler would always success when applied on PR
box and fail on the failure box
$P^{f}=\frac{1}{P(b'=1)}[\openone-P(b'=0)P^{\text{PR}}]$.

When $\xi=\frac{1}{\sqrt{2}}$, we could see a connection between QNC
and PQC in the transformation:
\begin{equation}
P^{\text{PR}}_{1/\sqrt{2}}(ab_{1}|xy_{1})P^{\text{PR}}_{1/\sqrt{2}}(b_{2}c|y_{2}z)\xrightarrow{\chi(b'=0)}P^{\text{PR}}_{1/2}(ac|xz).
\end{equation}
The final correlation are
nonlocal($\Vec{\text{CH}}\cdot\Vec{P}>\xi^{2}+1/2=1$) if and only if
the initial correlation is in the set of
PQC($\Vec{\text{CH}}\cdot\Vec{P}>\xi+1/2=1/\sqrt{2}+1/2$).

In this scheme, CH inequality is the measure of nonlocality. We find
that replacing the CH inequality with CHSH inequality\cite{CHSH} is
feasible for swapping joint probability-symmetric boxes, including
isotropic boxes. The characteristics are all tenable in the sense of
CHSH expression. The successful probability is
\begin{equation}
p(b'=0)=\frac{1}{6}\Vec{\text{CHSH}}\cdot\Vec{P}(b_{1}b_{2}|y_{1}y_{2})+\frac{1}{3}.
\end{equation}
Where
$\Vec{\text{CHSH}}\cdot\Vec{P}(ab|xy)=E_{xy}+E_{x\bar{y}}+E_{\bar{x}y}-E_{\bar{x}\bar{y}}$
with $E_{xy}=P(a=b|xy)-P(a\neq b|xy)$. The final correlation are
nonlocal ($\Vec{\text{CHSH}}\cdot\Vec{P}>4\xi^{2}=2$) if and only if
the initial correlation is in the set of PQC
($\Vec{\text{CHSH}}\cdot\Vec{P}>4\xi=2\sqrt{2}$).

\section{Generalized Svetlichny inequality for multipartite correlated boxes}\label{s3}
Before generalizing the swapping of bipartite nonlocalities to the
multipartite case, we need to make some theoretical preparations.

In tripartite system, a famous inequality for nonlocality is
Svetlichny\cite{Sv} inequality, and for nonsignaling box
$P(abc|xyz)$ it has the form\cite{resource}:
\begin{equation}
E_{xyz}+E_{xy\bar{z}}+E_{x\bar{y}z}-E_{x\bar{y}\bar{z}}+E_{\bar{x}yz}-E_{\bar{x}y\bar{z}}-E_{\bar{x}\bar{y}z}-E_{\bar{x}\bar{y}\bar{z}}\leq4,
\end{equation}
where $E_{xyz}$ is defined as
\begin{equation}
E_{xyz}=\sum_{a,b,c}(-1)^{a+b+c}P(abc|xyz).
\end{equation}
The box with probability distribution
\begin{equation}
P(abc|xyz)=
\begin{cases}
1/4,&\text{if }a\oplus b\oplus c=xy\oplus xz\oplus yz,\\
0,&\text{otherwise}.
\end{cases}
\end{equation}
gives the algebraic maximal violation 8, namely, the Svetlichny box
(SB), and the GHZ state
($\frac{1}{\sqrt{2}}(|000\rangle+|111\rangle$) achieves the quantum
maximal violation $4\sqrt{2}$ of Svetlichny inequality with an
appropriate set of measurements. From the joint probability
distribution matrix of GHZ we know that it is a linear superposition
of Svetlichny box and tripartite fully mixed box.

Now, we'll present the form of $n$-particle Svetlichny
inequality\cite{GSI} in nonsignaling models, namely, the generalized
Svetlichny inequality (GSI).

For a $n$-partite binary input and output correlated box
$P(a_{1}a_{2}\cdots a_{n}|x_{1}x_{2}\cdots x_{n})$, the expectation
values of measurements are defined as
\begin{equation}
\begin{aligned}
&E_{x_{1}x_{2}\cdots x_{n}}\\
&=\sum_{a_{1},a_{2},\cdots,a_{n}=0}^{1}(-1)^{\sum_{i=1}^{n}a_{i}}P(a_{1}a_{2}\cdots
a_{n}|x_{1}x_{2}\cdots x_{n}),
\end{aligned}
\end{equation}
and $E_{x_{1}x_{2}\cdots x_{n}}\in[0,1]$.

The generalized Svetlichny inequality has the form:
\begin{equation}
\text{GSI}^{n}=\sum_{x_{1},x_{2},\cdots,x_{n}=0}^{1}C_{x_{1},x_{2},\cdots,x_{n}}\cdot
E_{x_{1},x_{2},\cdots,x_{n}}\leq2^{n-1},
\end{equation}
The coefficients are defined as:
\begin{equation}
C_{x_{1},x_{2},\cdots,x_{n}}=\sqrt{2}\cos[\frac{\pi}{2}\cdot(\sum_{i=1}^{n}x_{i})\text{
mod }\emph{4}-\frac{\pi}{4}],
\end{equation}
where $X\text{ mod }n$ is addition modulo $n$ of $X$, and
$C_{x_{1},x_{2},\cdots,x_{n}}\in\{-1,1\}$.

When $n=2$, the GSI will be reduced to CHSH inequality and $n=3$ as
the Svetlichny inequality.

The generalized Svetlichny box (GSB) is characterized by the
following probability distribution:
\begin{equation}
\begin{aligned}
&P_{n}^{\text{GSB}}(a_{1},a_{2},\cdots,a_{n}|x_{1},x_{2},\cdots,x_{n})\\
&=
\begin{cases}
2^{1-n},&\text{if } \bigoplus_{i=1}^{n}a_{i}=\bigoplus_{j,k=1}^{n}x_{j}x_{k},\\
&\text{where }j\neq k\\
0,&\text{otherwise}.
\end{cases}
\end{aligned}
\end{equation}
Here the symbol $\bigoplus$ means addition modulo $\emph{2}$ of
summation. The generalized Svetlichny box violates the GSI up to its
algebraic maximum $2^{n}$.

If a box has the probability distribution
\begin{equation}
\begin{aligned}
&P_{n}(a_{1},a_{2},\cdots,a_{n}|x_{1},x_{2},\cdots,x_{n})\\
&=\prod_{i=1}^{n}P(a_{i}|x_{i})=[P(a_{i}|x_{i})]^{n}=(1/2)^{n},
\end{aligned}
\end{equation}
for all $a_{i}$ and $x_{i}$, it is the $n$-partite fully mixed box
$\openone^{n}(a_{1},a_{2},\cdots,a_{n}|x_{1},x_{2},\cdots,x_{n})$
(written as $\openone$ for simple later), and it has a violation of
zero in expression of GSI.

The $n$-partite isotropic boxes are defined as mixtures of
$n$-partite GSB and $n$-partite fully mixed box:
\begin{equation}
P_{\xi}^{\text{GSB}}=\xi P_{n}^{\text{GSB}}+(1-\xi)\openone,
\end{equation}
Where $\xi\in[-1,1]$. When $\xi=1/\sqrt{2}$, $\mathcal{B}_{\mathcal
{Q}}=\Vec{\text{GSI}}\cdot\Vec{P}_{1/\sqrt{2}}^{n}=2^{n-1/2}$ is
$n$-partite Tsirelson's bound, a generalized quantum bound. The
quantum state which can reach this bound is
${|\text{GHZ}\rangle}^{n}=\frac{1}{\sqrt{2}}(|0\rangle^{\otimes{n}}+|1\rangle^{\otimes{n}})$.

After defining the resources of multipartite nonlocality swapping,
we will begin the generalization of nonlocality swapping from the
two bipartite cases to the more general cases.

\section{Swapping two multipartite correlated boxes}\label{s4}
\subsection{Nonlocality swapping of two multipartite boxes}
In the previous scenario\cite{NLS}, Bob shares bipartite nonlocal
box with both Alice and Charlie. Nonlocal correlations will be
swapped with a certain probability when Bob applies the coupler
$\chi$ on his two boxes. Here we present a generalized nonlocality
swapping on two multipartite nonlocal correlated boxes.

\begin{figure}[htbp]
\includegraphics[width=0.45\textwidth]{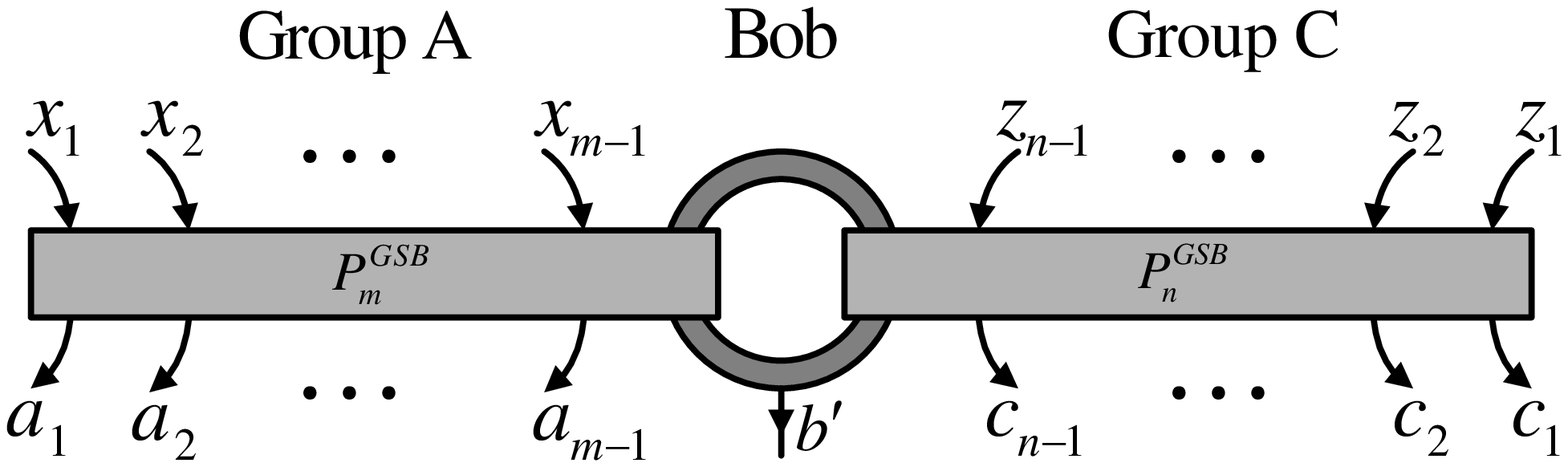}
\caption{\label{2m}In the nonlocality swapping scheme of two
multipartite boxes, Group A and B have $m$-partite and $n$-partite
nonlocal correlations respectively. Bob is the common user of two
groups. When he applies the coupler on his two boxes and gets a bit
0, the rest users of two groups will share an $m+n-2$-partite GSB.}
\end{figure}

Bob applies the bipartite coupler $\chi$ on two GSBs and swaps the
correlations between two groups of users (FIG.\ref{2m}). Besides
Bob, there are $m-1$ and $n-1$ users in group A and C, respectively,
which have inputs
$x_{1}x_{2}{\cdots}x_{m-1},z_{1}z_{2}{\cdots}z_{n-1}$ and outputs
$a_{1}a_{2}{\cdots}a_{m-1},c_{1}c_{2}{\cdots}c_{n-1}$. The coupler
acting on two GSBs implements a linear transformation:
\begin{widetext}
\begin{equation}
\begin{aligned}
&P_{m}^{\text{GSB}}(a_{1}a_{2}{\cdots}a_{m-1}b_{1}|x_{1}x_{2}{\cdots}x_{m-1}y_{1})P_{n}^{\text{GSB}}(b_{2}c_{1}c_{2}{\cdots}c_{n-1}|y_{2}z_{1}z_{2}{\cdots}z_{n-1})\\
&\xrightarrow{\chi}P(a_{1}a_{2}{\cdots}a_{m-1}b'c_{1}c_{2}{\cdots}c_{n-1}|x_{1}x_{2}{\cdots}x_{m-1}z_{1}z_{2}{\cdots}z_{n-1})\\
&=
\begin{cases}
p(b'=0)P_{m+n-2}^{\text{GSB}}(a_{1}a_{2}{\cdots}a_{m-1}c_{1}c_{2}{\cdots}c_{n-1}|x_{1}x_{2}{\cdots}x_{m-1}z_{1}z_{2}{\cdots}z_{n-1}),\\
p(b'=1)P^{f}(a_{1}a_{2}{\cdots}a_{m-1}c_{1}c_{2}{\cdots}c_{n-1}|x_{1}x_{2}{\cdots}x_{m-1}z_{1}z_{2}{\cdots}z_{n-1}).
\end{cases}
\end{aligned}
\end{equation}
\end{widetext}
The coupler encompasses the inputs and outputs of Bob's boxes and
returns a single bit $b'$, Bob succeeds in swapping a generalized
Svetlichny correlation between Group A and Group C with probability
$p(b'=0)=\frac{1}{6}\Vec{\text{CHSH}}\cdot\Vec{P}(b_{1}b_{2}|y_{1}y_{2})+\frac{1}{3}$,
and then an $m+n-2$-partite GSB is generated. If both $m$ and $n$
are greater than $2$, the final correlation is among more users than
the initial ones ($m+n-2$ is greater than both $m$ and $n$).

\subsection{Emerge of quantum bound}
After swapping the extremal nonlocal boxes, we'll consider the
swapping of noisy correlations. Group A and C now have the
multipartite isotropic boxes $P^{\text{GSB}}_{m,\xi}=\xi
P^{\text{GSB}}_{m}+(1-\xi)\openone$ and $P^{\text{GSB}}_{n,\xi}=\xi
P^{\text{GSB}}_{n}+(1-\xi)\openone$ instead, and Bob will apply the
coupler on his two boxes to check whether the quantum bound emerges
or not. Fortunately, the answer is positive. the coupling process
will be like this:
\begin{equation}
P^{\text{GSB}}_{m,\xi}P^{\text{GSB}}_{n,\xi}\xrightarrow{\chi(b'=0)}P^{\text{GSB}}_{m+n-2,\xi^2}=\xi^2P^{\text{GSB}}_{m+n-2}+(1-\xi^2)\openone.
\end{equation}
By setting the coefficient $\xi=1/\sqrt{2}$, we can get the result
that only the postquantum type initial correlations can make the
final box nonlocal, which shows the generalized Tsirelson's bound
$\mathcal{B}_{\mathcal {Q}}=2^{(n'-1/2)}$ (here $n'=m+n-2$) again.

\section{Swapping many multipartite nonlocality}\label{s5}
\subsection{Multipartite nonlocality swapping}
After presenting the generalization in section IV, we now introduce
a more general nonlocality swapping scheme. Bob shares nonlocal
boxes with three or more groups of users, and his new purpose is
swapping nonlocal correlation between all of the groups. Toward this
new goal, Bob makes a feasible try by using bipartite coupler
$\chi$. He will show the probability to swap nonlocal correlation
between groups of users by demonstrating the simplest example where
he shares PR boxes with other three users Alice(A), Charlie(C) and
Danny(D).

\begin{figure}[htbp]
\includegraphics[width=0.40\textwidth]{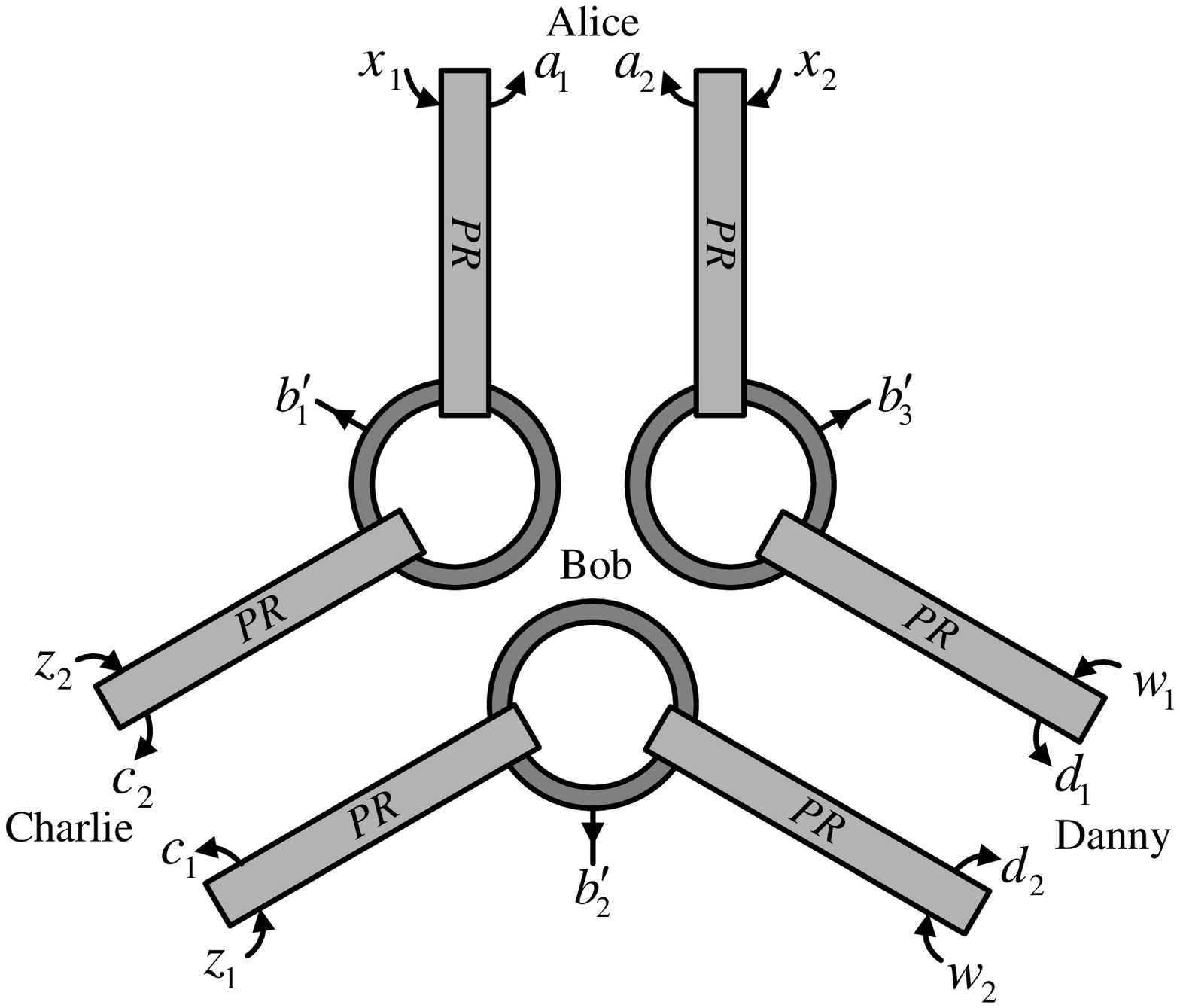}
\caption{\label{62}A hybrid scheme for swapping the bipartite PR
correlations to tripartite Svetlichny correlation. Bob applies three
couplers on six PR boxes, and gets three output bits $b'_{1}$,
$b'_{2}$ and $b'_{3}$. Alice, Charlie and Danny applies the local
operation: inputs $x_{1}=x_{2}=x$, $z_{1}=z_{2}=z$, $w_{1}=w_{2}=w$
and outputs $a=a_{1}{\oplus}a_{2}$, $c=c_{1}{\oplus}c_{2}$,
$d=d_{1}{\oplus}d_{2}$. If $b'_{1}=b'_{2}=b'_{3}=0$, a Svetlichny
box is generated from the initial PR boxes.}
\end{figure}

As depicted in FIG.\ref{62}, Bob has two copies of PR box with A, C
and D each. First, Bob applies the coupler on the boxes he shares
with A and C, C and D, A and D separately, and then he will get
three outputs after his actions. The process succeeds if the three
outputs are all bit $0$. Second, some corresponding local operations
(LO) must be applied on A, C and D's boxes. After that, an
Svetlichny correlation is generated between A, C and D.

Suppose that Bob's correlation $P(b_{1}b_{2}b_{3}|y_{1}y_{2}y_{3})$
is fully mixed correlation, and the transformation of this scheme
should be written as:
\begin{equation}
\begin{aligned}
&\left.\begin{aligned}
P^{\text{PR}}(a_{1}b_{1}|x_{1}y_{1})P^{\text{PR}}(c_{2}b_{2}|z_{2}y_{2})\xrightarrow{\chi,LO}\\
P^{\text{PR}}(c_{1}b_{3}|z_{1}y_{3})P^{\text{PR}}(d_{2}b_{4}|w_{2}y_{4})\xrightarrow{\chi,LO}\\
P^{\text{PR}}(d_{1}b_{5}|w_{1}y_{5})P^{\text{PR}}(a_{2}b_{6}|x_{2}y_{6})\xrightarrow{\chi,LO}\\
\end{aligned}
\right\rbrace P(ab'_{1}b'_{2}b'_{3}cd|xzw)\\
&=
\begin{cases}
\frac{1}{27}P^{\text{SB}},&\text{if }b'_{1}=b'_{2}=b'_{3}=0,\\
\frac{6}{27}(\frac{3}{2}\openone-\frac{1}{2}P^{\text{SB}}),&\text{if two of }b'_{1},b'_{2},b'_{3}\text{ equal }0,\\
\frac{12}{27}(\frac{3}{4}\openone+\frac{1}{4}P^{\text{SB}}),&\text{if two of }b'_{1},b'_{2},b'_{3}\text{ equal }1,\\
\frac{8}{27}(\frac{9}{8}\openone-\frac{1}{8}P^{\text{SB}}),&\text{if }b'_{1}=b'_{2}=b'_{3}=1.\\
\end{cases}
\end{aligned}
\end{equation}

\emph{Analysis of the scheme}: In this scheme, Bob realizes a
nonlocality swapping scheme, which generates a tripartite nonlocal
extremal correlation from initial six bipartite extremal ones.
Because three bipartite couplers are applied, the probability of
successful swapping is only $1/27$, and it seems too small. This
method can be extended to generate  multipartite nonlocal
correlation from sufficient bipartite ones. $N(N-1)/2$ couplers will
act on $N(N-1)$ PR boxes and generate a $N$-partite GSB with
probility $1/3^{(N(N-1)/2)}$. In essence, it
is a hybrid scheme of Skrzypczyk's scheme and simulating tripartite
boxes\cite{resource}, and obviously, an less efficient and high
consumption scheme. So, we will give another more efficient and
low-consumption swapping scheme for many multipartite boxes.

\subsection{Swapping many boxes by multipartite coupler $\chi_{N}$}
We will present our generalization, a more efficient scheme for
multipartite nonlocality swapping than the hybrid one. In this
scheme, only one generalized multipartite coupler is needed, and the
probability of success is the same as Skrzypczyk's scheme.

\begin{figure}[htbp]
\includegraphics[width=0.45\textwidth]{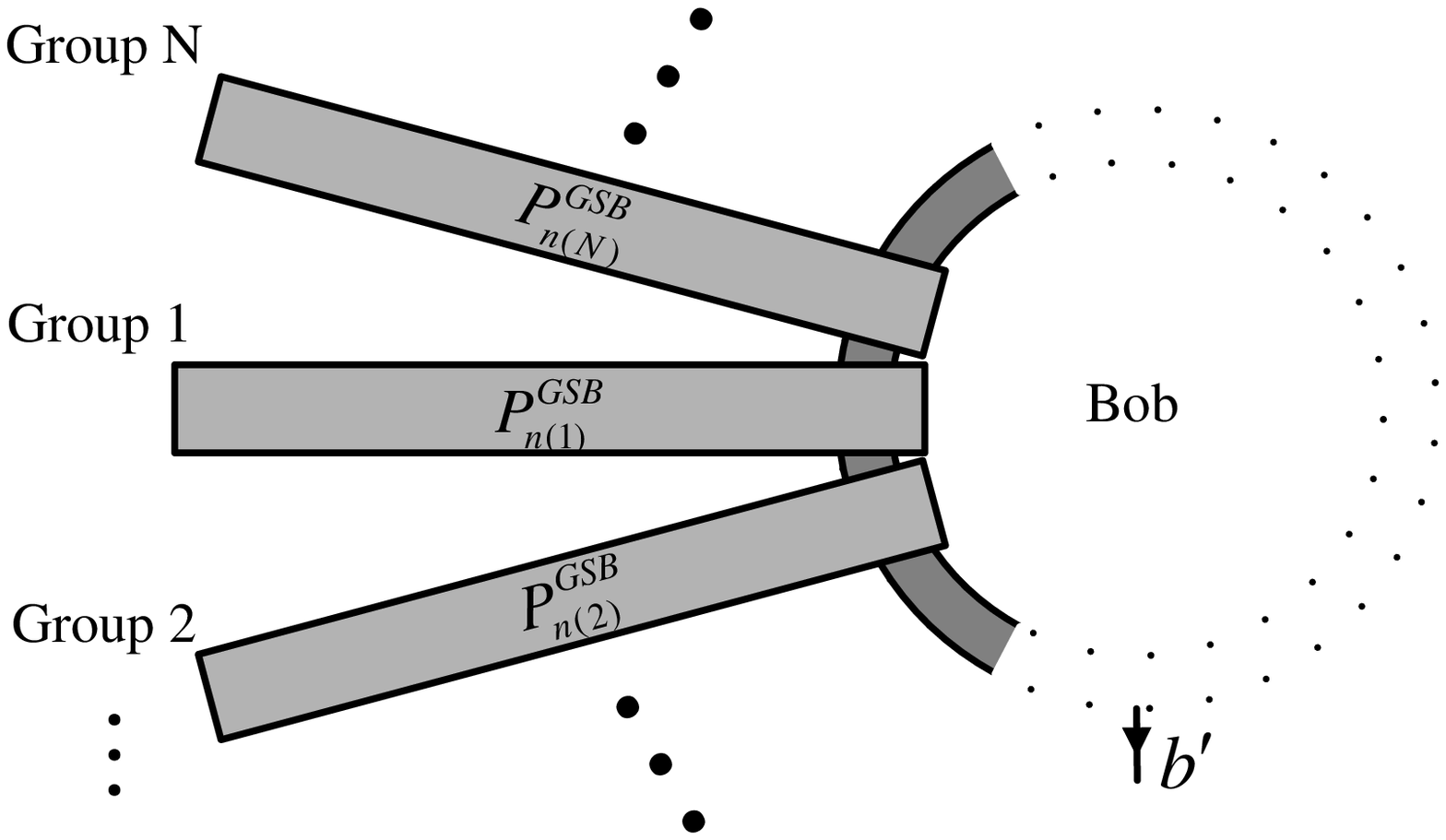}
\caption{\label{mm}Multipartite coupler for many boxes. Bob applies
the coupler on $N$ GSBs and gets a bit $b'$. The generalized coupler
is an analogue of quantum joint measurement on N particles, among
which each comes from a multipartite entangled state in
multiparticle entanglement swapping.}
\end{figure}

\emph{Generalized multipartite coupler $\chi_{N}$}: In the protocol
of multipartite entanglement swapping\cite{multiES}, a joint
measurement is performed on N particles, among which each comes from
a multipartite entangled state, and then a final multipartite
entangled state is generated between the rest users. Our generalized
coupler applying on many boxes is the analogue of this joint
measurement on N particles(see in FIG.\ref{mm}). The coupler
$\chi_{N}$ will return a single bit $b'$ when applied on $N$
nonlocal boxes, which will be the indication of whether the coupling
is success. The probability of output bit $b'=0$ is a linear
function about the correlations of boxes that the coupler is acting
on.

\emph{The generalization of nonlocality swapping}: Now, Bob get a
new device-the multipartite coupler $\chi_{N}$, and he will try to
swap nonlocal correlation between many multipartite correlated boxes
again.

There are $N$ groups of users and the $i$th ($i=1,2,\cdots,N$) group
shares a $n(i)$-partite GSB. Bob is a special user who belongs to
every group, and he will carry out the generalized coupler on all
boxes. The transformation is defined as follows:
\begin{equation}
\begin{aligned}
&P_{n(1)}^{\text{GSB}}(A_{1}b_{1}|X_{1}y_{1})P_{n(2)}^{\text{GSB}}(A_{2}b_{2}|X_{2}y_{2}){\cdots}P_{n(N)}^{\text{GSB}}(A_{N}b_{N}|X_{N}y_{N})\\
&\xrightarrow{\chi_{N}}
\begin{cases}
p(b'=0)P_{{\sum}n(i)-N}^{\text{GSB}}(A|X),\\
p(b'=1)P^{f}(A|X).\\
\end{cases}
\end{aligned}
\end{equation}
Where $A_{i}$ means the $i$th box's all outputs
$a_{1}a_{2}{\cdots}a_{n(i)-1}$ except Bob's $b_{i}$, and $X_{i}$
means the $i$th box's inputs $x_{1}x_{2}{\cdots}x_{n(i)-1}$ except
Bob's $y_{i}$. The $A$ and $X$ without subscript mean all users'
input bits and output bits except Bob's. One
$\sum_{i=1}^{N}n(i)-N$-partite correlation will be generated after a
successful coupling in Bob's location.

\begin{figure}[htbp]
\includegraphics[width=0.35\textwidth]{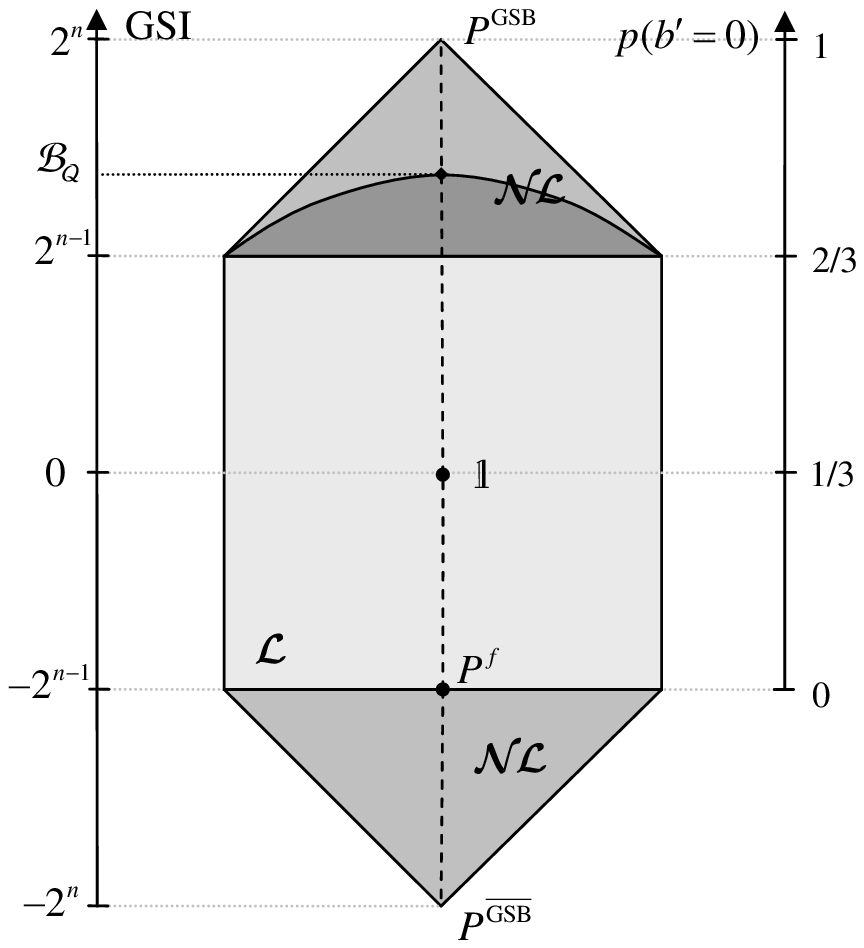}
\caption{\label{set}The set of resources for the generalized
nonlocality swapping. The left axis is the violation under GSI
expression. Local states satisfy
$-2^{n-1}\leq\Vec{\text{GSI}}\cdot\Vec{P}^{L}\leq2^{n-1}$, the other
states have nonlocal correlation otherwise. Generalized Tsirelson's
bound $\mathcal {B}_{\mathcal {Q}}=2^{n-1/2}$ lies on the
multipartite isotropic boxes (the dashed line). The right axis is
the success probability of the generalized coupler. The coupler is
only allowed to be applied on the area:
$-2^{n-1}\leq\Vec{\text{GSI}}\cdot\Vec{P}\leq2^{n}$. Note the
polytope of multipartite resource is a high-dimensional system,
while the figure is a two-dimensional illustration.}
\end{figure}

Suppose that a perfect coupler works on the region of nonsignaling
polytope depicted in FIG.\ref{set}, and then the rest users will get
a local correlation which violates the GSI by $-2^{n-1}$ when
swapping fails; a GSB will be generated when the coupler returns a
single bit $0$. The output is deterministic $1$ when applying the
coupler on the failure correlation $P^{f}(A|X)$ and $0$ on
generalized Svetlichny correlation. Then the generalized coupler
acting on any allowed box gives a successful probability:
\begin{equation}
p(b'=0)=\frac{1}{3\cdot2^{n-1}}\Vec{\text{GSI}}\cdot\Vec{P}+\frac{1}{3}.
\end{equation}
For a natural correlation $\openone$, the optimal probability of
success is also $1/3$ as before.

\emph{Comments on the generalization}: Nonlocality swapping makes it
possible to generate multipartite nonlocal correlation from many
nonlocal correlations. For instance, we could generate a $4$-partite
Svetlichny box from one tripartite Svetlichny box and two PR boxes
by a generalized coupler $\chi_{3}$.

The generalized swapping process is also a linear transformation.
Let $N=2$, the scheme becomes the one we presented in Sec. IV. And
further, let $n(1)=n(2)=2$, it is the same as Skrzypczyk's scheme.

When the noisy condition is considered, Tsirelson's bound for
quantum nonlocality doesn't always appear in the swapping process.
Basing on the linearity of the swapping, if all group's correlations
are noisy like
$P_{\xi}^{\text{GSB}}={\xi}P_{n(i)}^{\text{GSB}}+(1-\xi)\openone$,
the final correlation will be
$P_{\xi^{N}}^{\text{GSB}}={\xi^{N}}P_{{\sum}n(i)-N}^{\text{GSB}}+(1-\xi^{N})\openone$.
The quantum bound only appears in the case where two boxes are noisy
or the coupler is $\chi_{2}$.

\section{Conclusions}\label{s6}
We found that CHSH inequality is also pragmatic for previous
nonlocality swapping scheme. Then we focused on the Svetlichny
inequality in tripartite system, and showed the form of $N$-Particle
Svetlichny inequality in multipartite nonsignaling system. Basing on
this generalized Bell-type inequality, we defined the extremal
multipartite nonlocal boxes for our generalization of swapping
process in the paper.

Using the same coupler applied in Skrzypczyk's swapping scheme, we
first presented a generalized swapping of two arbitrary multipartite
nonlocal correlations, which could generate a multipartite
correlation with more users than the initial ones. Later, through a
combined scheme, we illustrated that multipartite correlation could
also be generated from swapping many bipartite boxes. For the sake
of efficiency, we finally presented a more general multipartite
nonlocality swapping scheme with a generalized multipartite coupler,
an analogue of quantum joint measurement in multipartite
entanglement swapping. The multipartite coupler builds an efficient
device to generate multipartite nonlocal correlation between many
users.

The generalized quantum bound always emerges in the scheme of
swapping two multipartite isotropic boxes, and occasionally emerges
in the process of swapping three or more multipartite boxes. Judged
by appearance, the emergence of quantum bound is merely a numerical
coincidence, but the mathematical relation between nonlocal
criterion ($2^{n-1}$), quantum bound ($2^{n-1/2}$) and extremal
violation ($2^{n}$) is a clue to get a deep understanding of quantum
nonlocality.

Many open problems are presented in front of us. The fact that
nonlocal correlation only can be generated from swapping the
postquantum boxes makes not only an emergence of quantum bound but
also a gap between QNC and PQC. The coupler, as the most important
part in swapping process, beyond the device for poor dynamics
applied outside the nosigned boxes after measurement, is very likely
valuable for some other dynamical processes, individually or
combined as the scheme depicted in FIG.\ref{62}. How to generalize the
swapping to high dimensional systems (multipartite multi-nary boxes)
with suitable inequality is worth to study, too.

\begin{acknowledgments}
This work is supported by National Natural Science Foundation of
China (NSFC) under Grants No. 10704001, No. 61073048 and 11005029,
the Key Project of Chinese Ministry of Education.(No.210092),
the China Postdoctoral Science Foundation under Grant No. 20110490825, the
Key Program of the Education Department of Anhui Province under
Grants No. KJ2008A28ZC, No. 2010SQRL153ZD, and No. KJ2010A287, the
`211' Project of Anhui University, the Talent Foundation of Anhui
University under Grant No.33190019, the personnel department of
Anhui province, and Anhui Key Laboratory of Information Materials
and Devices (Anhui University).
\end{acknowledgments}

\end{document}